# Crystal field potential and short-range order effects in inelastic neutron scattering, magnetization and heat capacity of the cage-glass compound HoB$_{12}$

B. Z. Malkin[a], E. A. Goremychkin[b], K. Siemensmeyer[c], S. Gabáni[d], K. Flachbart[d], M. Rajňák[d], A. L. Khoroshilov[e], K. M. Krasikov[e], N. Yu. Shitsevalova[g], V. B. Filipov[g], N. E. Sluchanko*[e]

[a]Kazan Federal University, Kazan, 420008 Russia
[b]Frank Laboratory of Neutron Physics, Joint Institute for Nuclear Research, Joliot-Curie 6, Dubna, 141980 Moscow reg., Russia
[c]Helmholtz Zentrum Berlin, D 14109 Berlin, Germany
[d]Institute of Experimental Physics SAS, Watsonova 47, 04001 Košice, Slovakia
[e]Prokhorov General Physics Institute of Russian Academy of Sciences, Vavilova 38, 119991 Moscow, Russia
[g]Frantsevich Institute for Problems of Materials Science, NASU, Krzhyzhanovsky str., 3, 03142 Kyiv, Ukraine

**Abstract**

The strongly correlated system Ho$^{11}$B$_{12}$ with boron sublattice Jahn-Teller instability and nanoscale electronic phase separation (dynamic charge stripes) was studied in detail by inelastic neutron scattering (INS), magnetometry and heat capacity measurements at temperatures in the range 3-300 K. From the analysis of registered INS spectra, we determined parameters of the cubic crystal field at holmium sites, $B_4$= - 0.333 meV and $B_6$= -2.003 meV (in Stevens notations), with an unconventional large ratio $B_6/B_4$ pointing on the dominant role of conduction electrons in the formation of a crystal field potential. The molecular field in the antiferromagnetic state, $B_{loc}$ = (1.75± 0.1) T has been directly determined from the INS spectra together with short-range order effects detected in the paramagnetic state. A comparison of measured magnetization in diluted Lu$_{0.99}$Ho$_{0.01}$B$_{12}$ and concentrated HoB$_{12}$ single crystals showed a strong suppression of Ho magnetic moments by antiferromagnetic exchange interactions in holmium dodecaboride. To account explicitly for the short-range antiferromagnetic correlations, a self-consistent holmium dimer model was developed that allowed us to reproduce successfully field and temperature variations of the magnetization and heat capacity in the cage-glass phase of HoB$_{12}$ in external magnetic fields.

## I. INTRODUCTION

Rare earth (RE) dodecaborides $RB_{12}$ (*fcc* lattice, space group Fm3m, R = Tb, Dy, Ho. Er, Tm, Yb and Lu) with a cage-glass structure [1] attract considerable attention due to a unique combination of their physical properties, including high melting temperature, microhardness, high chemical stability, etc. In the antiferromagnetic (AF) $RB_{12}$ series the Néel temperature decreases monotonically from $TbB_{12}$ ($T_N \approx 22$ K) to $TmB_{12}$ ($T_N \approx 3.2$ K) while maintaining similar conduction band, consisting of $5d$ (R) and $2p$ (B) atomic orbitals, and changing only with the number $n_{4f}$ of $4f$ shell electrons of RE ions ($8 \leq n_{4f} \leq 14$) [2,3]. The increase of the $4f$ filling from $n_{4f}=12$ to $n_{4f}=13$ leads to dramatic changes both in magnetic and charge transport characteristics [4-6], demonstrating the transition from AF metal ($TmB_{12}$) to paramagnetic narrow-gap semiconductor with intermediate valence [7] of Yb ions ($YbB_{12}$). In spite of intensive investigations, the nature of this semiconducting state in Kondo insulator $YbB_{12}$ remains a subject of active debates [8-13].

In recent studies of the charge transport, magnetic and thermal properties and fine details of *fcc* crystal structure of non-magnetic $LuB_{12}$, antiferromagnetic $HoB_{12}$-$TmB_{12}$ compounds and solid solutions $Tm_{1-x}Yb_xB_{12}$, it was established that the cooperative Jahn-Teller dynamics of $B_{12}$ clusters should be considered as one of the main factors responsible for a strong renormalization of the quasiparticle spectra, electron phase separation and the symmetry breaking in RE dodecaborides [14-18]. It was suggested that the ferrodistortive effect in the boron sublattice generates both collective modes (overdamped oscillators in the frequency range 250-1000 cm$^{-1}$ [17-18] in the dynamic conductivity spectra of each of the metallic $RB_{12}$ compounds and rattling modes, - quasi-local vibrations of heavy rare earth ions embedded in the oversized $B_{24}$ cavities. Large amplitude displacements of R-ions cause (*i*) the development of vibrational instability at intermediate temperatures ~150 K on reaching the Ioffe-Regel limit for a localization and (*ii*) strong periodic changes in the hybridization of atomic $5d$ (R) and $2p$ (B) orbitals, altering their overlap. Accordingly, these overlap alterations along the rhombic (of [110]-type) axes are responsible for the modulation of the conduction electron density with a frequency $\sim 2 \cdot 10^{11}$ Hz [17] providing the emergence of dynamic charge stripes which are the feature of a nanoscale electron instability and electron phase separation. It was argued in [17-18] that non-equilibrium charge carriers dominated in RE dodecaborides taking 50-70% from the total number of conduction electrons. These *hot electrons* cannot participate in the indirect exchange (RKKY interaction) between the RE magnetic moments. Moreover, it was concluded in the studies of $HoB_{12}$ and $Ho_xLu_{1-x}B_{12}$ that the suppression of the magnetic exchange interaction in nano-size filamentary structures should be considered as the main sequence of the electron instability providing a symmetry lowering and formation of Maltese cross-like phase diagrams of $RB_{12}$ antiferromagnets [19-20]. Because of the loosely bound state of RE ions in the rigid boron cage, there is also an order-disorder phase transition to the cage-glass state [1] at temperature $T^*\sim 60$ K resulting in the random displacements of RE ions from their positions in the *fcc* lattice. This disordering induces a formation of nano-size clusters of RE-ions (dimers, trimers, etc.) with AF exchange, and corresponding domains (which in $HoB_{12}$ have a form of a cigar elongated parallel to trigonal axes (of [111]-type) [21]) should be taken into account to explain the nature of the short-range magnetic order detected in $RB_{12}$ antiferromagnets at temperatures far above $T_N$.

As with all rare earth systems, the interaction of *f*-electrons with the crystal field (CF) plays a key role in determining magnetic and thermodynamic properties of RE dodecaborides. Taking into account that the CF potential is sensitive to structural distortions and magnetic ions displacement as well as to changes in the conduction electron density, it looks useful to investigate in detail the *4f* shell ground multiplet splitting in magnetic dodecaborides. Inelastic neutron scattering (INS) is the most reliable method of determining the CF potential in metallic compounds since it measures directly energies of CF excitations and intensities of magnetic dipole transitions between CF levels. In the present study, samples of $HoB_{12}$ were used to characterize the CF potential and short-range order effects in the dodecaboride matrix by INS, magnetization and heat capacity measurements.

## II. EXPERIMENTAL DETAILS

Since the natural mixture of boron isotopes has a high neutron absorption cross-section (767 barn), samples for INS experiments were prepared using the $^{11}B$ isotope whose absorption cross-section is nearly zero (0.006 barn). Along with $Ho^{11}B_{12}$, an isostructural compound $Lu^{11}B_{12}$ was also synthesized. Since lutetium has fully filled the *4f* electronic shell and does not have a magnetic moment, the INS measurement with $Lu^{11}B_{12}$ makes it possible to estimate the contribution of nuclear (phonon) scattering and to establish the nature of features (magnetic/nonmagnetic) in the INS spectra of $Ho^{11}B_{12}$.

The powder samples of $Ho^{11}B_{12}$ and $Lu^{11}B_{12}$ were synthesized in two stage process. Firstly, the initial charges were prepared as a mixture of corresponding metal oxide ($Ln_2O_3$) and isotopic boron ($^{11}B$) for subsequent synthesis according to the solid-state reaction

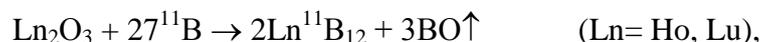
$$Ln_2O_3 + 27\,^{11}B \rightarrow 2Ln^{11}B_{12} + 3BO\uparrow \qquad (Ln= Ho, Lu),$$

each mixture was pressed in the form of tablets and then was held in vacuum at 1900 K during 1 hour. The tablets of synthesized dodecaborides $Ln^{11}B_{12}$ were ground, then newly pressed and held at 2000 K for homogenization. Further the tablets were ground in a steel mortar, and the obtained powders were treated by hot HCl acid to remove traces of Fe. X-ray phase analysis showed that the prepared powders of Ho$^{11}$B$_{12}$ and Lu$^{11}$B$_{12}$ are single-phase products of UB$_{12}$-type *fcc* crystal structure. These Ln$^{11}$B$_{12}$ powders were prepared using isotope-enriched boron ($^{11}$B) (Ceradyne Inc., USA, with enrichment of 99.61 at. %, purity >99.96 %) and metal oxides from the Federal State Research and Design Institute of Rare Metal Industry (Moscow, Russia) with the purity of initial Ho$_2$O$_3$ of 99.996 and Lu$_2$O$_3$ of 99.998 mass %. Neutron diffraction measurements confirmed that the samples contained no other phases.

The INS experiments were performed at the pulsed reactor IBR-2 (FLNP, JINR, Dubna, Russia) on the time-of-flight inverse geometry spectrometer NERA [22], with a final energy of 4.65 meV and energy resolution at the elastic position of 0.8 meV, and on the direct geometry spectrometer DIN 2PI [23], using incident energy of 5 meV with a resolution of 0.28 meV. The samples, weighting about 12 g, were placed in thin aluminium sachet and inserted into a closed-cycle cryostat for scans in the temperature range 3-300 K.

Measurements of the magnetization field dependences for the oriented single crystalline Ho$_{0.01}$Lu$_{0.99}$B$_{12}$ samples in magnetic fields up to 18 T and HoB$_{12}$ in the paramagnetic phase in external magnetic fields up to 7 T directed along the tetragonal, rhombic and trigonal symmetry axes were performed using vibrating sample magnetometer with cryogen-free 18 T magnet (Cryogenic Limited, UK) and SQUID MPMS-7 (Quantum Design, USA) instrument, correspondingly.

Precise measurements of the heat capacity of HoB$_{12}$ and LuB$_{12}$ single crystalline samples at temperatures in the range 1.8–300 K in magnetic fields up to 9 T directed along [001], [110] and [111] axes were performed using PPMS-9 (Quantum Design) instrument. The details of the growth of Ho$_{0.01}$Lu$_{0.99}$B$_{12}$, Ho$^{11}$B$_{12}$, and LuB$_{12}$ single crystals and the preparation of oriented samples are described in [1,20].

### III. INS RESULTS

Figure 1 shows the results of the INS measurement on NERA spectrometer at temperature of 45 K for the compounds Ho$^{11}$B$_{12}$ and Lu$^{11}$B$_{12}$. Comparison of Ho$^{11}$B$_{12}$ and Lu$^{11}$B$_{12}$ spectra demonstrates the presence of five well-defined magnetic peaks (marked by vertical arrows) due to transitions between CF levels at energy transfers of $E_1 \approx 2.3$ meV, $E_2 \approx 7.2$ meV, $E_3 \approx 9.5$ meV, $E_4 \approx 17.4$ meV and $E_5 \approx 26.7$ meV. The temperature evolution of the scattering intensities shown in Fig. 2 indicates that the intensity of peaks $E_2$, $E_3$, $E_4$ and $E_5$ decreases with increasing temperature, which is a signature of transitions from the ground CF level, while the $E_1$ peak intensity increases with increasing temperature and therefore this peak is associated with the transition from the excited CF level ($E_2$) to the next level ($E_3$).

Figure 3 shows the results on Ho$^{11}$B$_{12}$ in the antiferromagnetic (AF) state at $T$=3.5 K well below the Néel temperature $T_N$=7.4 K [20,21]. The data indicate the appearance of a new peak at energy $E_{AF}$= 0.8 meV and a significant shift in the energy position of peaks $E_2$ and $E_3$, as well as a change in their intensities. This indicates a strong influence of the magnetic order on the energy and intensity of CF transitions. It is worth noting that the $E_{AF}$ value is well comparable with the doubled spin gap energy $2\Delta_g \approx 0.8$ meV detected at 3.5 K in the magnon excitation spectra of HoB$_{12}$ [21].

## IV. INS DATA ANALYSIS AND CF MODEL

In the case of perfect *fcc* UB$_{12}$-type lattice of Ho$^{11}$B$_{12}$ it is expected that holmium ions occupy sites with cubic point symmetry. Neglecting interactions between Ho$^{3+}$ ions, in particular, in the paramagnetic phase, we write the Ho$^{3+}$ single-ion Hamiltonian operating in the total space of 1001 states of the 4$f^{10}$ electronic configuration in the following form

$$H = H_{FI} + H_Z + H_{CF} \tag{1}$$

where $H_{FI}$ is the free ion Hamiltonian considered in the standard parameterized form [24] with parameters taken from Ref. 25, $H_Z = -\mathbf{M} \cdot \mathbf{B}$ is the electronic Zeeman energy in external magnetic field $\mathbf{B}$ (here $\mathbf{M} = -\mu_B \sum_n (k\mathbf{l}_n + 2\mathbf{s}_n)$ is the magnetic moment operator of a Ho$^{3+}$ ion, $\mu_B$ is the Bohr magneton, the sum is taken over 4$f$ electrons with orbital $\mathbf{l}_n$ and spin $\mathbf{s}_n$ moments, respectively, $k$ is the orbital reduction factor), and

$$H_{CF} = 8B_4[C_0^{(4)} + (5/14)^{1/2}(C_4^{(4)} + C_{-4}^{(4)})] + 16B_6[C_0^{(6)} - (7/2)^{1/2}(C_4^{(6)} + C_{-4}^{(6)})] \tag{2}$$

is the CF Hamiltonian determined by phenomenological CF parameters $B_4$ and $B_6$ (here $C_n^{(m)}$ are spherical tensor operators which can be transformed to Stevens operators $O_m^n(\mathbf{J})$ with the reduced matrix elements $\alpha_J, \beta_J, \gamma_J$ in a truncated basis of eigenfunctions $|J, J_z>$ of the angular momentum $J$, in particular, $C_0^{(4)} = \beta_J O_4^0 / 8$ and $C_0^{(6)} = \gamma_J O_6^0 / 16$ [26]). When studying scattering of slow neutrons, magnetic and thermal properties of a holmium compound, we may limit ourselves by considering only the ground multiplet $^5I_8$ of the 4$f^{10}$ electronic shell (the first excited multiplet $^5I_7$ has an energy exceeding 7000 K). The $^5I_8$ multiplet is split by CF of cubic symmetry into seven sublevels,- four triplets, two doublets, and one singlet.

The intensity of INS in a powder sample at temperature $T$ corresponding to magnetic dipole transitions between CF levels with the energy transfer $\hbar\omega$ and the scattering vector $\mathbf{Q}$ is given by the expression [27]:

$$S(Q,\omega) \propto |f(Q)|^2 \sum_{i,j>i} \rho_i(T) \sum_{\alpha=x,y,z} |<j|M_\alpha|i>|^2 F(\omega-\omega_{ji},\Gamma_{ji}) \qquad (3)$$

where $f(Q)$ is the $f$-electron form factor, $\rho_i(T) = \exp(-E_i/k_BT)/\sum_j \exp(-E_j/k_BT)$ (here $k_B$ is the Boltzman constant) is the occupation probability of the CF level $|i\rangle$ with the energy $E_i$, and $F(\omega-\omega_{ji},\Gamma_{ji})$ is the line shape with a half-width $\Gamma_{ji}$. In the present work the Lorentzian function has been used.

In our INS data we observed two strong transitions at the energies near $E_2$=7.2 meV and $E_3$=9.5 meV. The temperature evolution of their intensities indicates that they are due to transitions between the ground state and the first and second excited CF levels, correspondingly. From the fitting of observed scattering profiles under the condition of $E_3/E_2 \approx 1.32$ and by making use of the setup resolution of 0.8 meV as a starting value of the half-width $\Gamma_{ji}$ for all transitions, we obtained CF parameters $B_4$ = (-0.333 ± 0.012) meV and $B_6$ = (-2.003 ± 0.023) meV. These values of CF parameters allowed us to reproduce successfully the experimental INS spectra at 15 K, 45 K, 100 K and 230 K, solid lines in Fig. 2 represent the results of the profile refinements of magnetic scattering using slightly varied half-widths (mainly increasing with the energy transfer) for different transitions. It can be seen that the model describes well the evolution with temperature of all the transitions. There are a few minor discrepancies between the calculated and measured spectra although the overall agreement is very good for the CF-only model. The calculated energies of CF sublevels of the ground multiplet and corresponding irreducible representations of the point symmetry $O_h$ group are presented in Table I. The energy spectrum of the ground state, the $\Gamma_5$ triplet, in an external magnetic field $\mathbf{B}$ can be described by the effective Spin-Hamiltonian ($S$=1) with the isotropic g-factor $H_S = g\mu_B \mathbf{S}\cdot\mathbf{B}$.

Holmium has only one stable isotope $^{165}$Ho with the nuclear spin $I$=7/2. By making use the numerical diagonalization of the projection of the Hamiltonian (1) on the space of 136 electron-nuclear states $|J,I,J_z,I_z>$ of the ground multiplet with added magnetic dipole hyperfine interaction $H_{HF} = A\mathbf{J}\cdot\mathbf{I}$ (here the hyperfine coupling constant $A$=3.3·10$^{-3}$ meV [28]), we calculated the widths $\Delta_{HF}$ of hyperfine structures of CF sublevels presented in Table I. It is seen that these widths are remarkably smaller than the INS spectrometers resolution, and we obtain practically the same scattering profiles from calculations where the hyperfine interaction is added to the single-ion Hamiltonian.

Additionally, from the data obtained, we can conclude about slight energy changes with temperature of the $E_2$ and $E_3$ peaks. Indeed, within the limits of experimental accuracy, the increase of $E_2(T)$ and $E_3(T)$ with decreasing temperature is well evident below 30 K (see inset in Fig. 3). It is worth noting that in recent measurements of electron paramagnetic resonance (EPR) in $Ho_xLu_{1-x}B_{12}$ [29] strong and simultaneous changes of both the *g*-factor (*g*=4.1-4.8) and the line width ($\Delta B$=0.35-0.9 T) have been observed in $HoB_{12}$ at temperatures between 7.5 and 30 K. This result was discussed by the authors of [29] in terms of short-range AF clusters which were observed in the paramagnetic state of $HoB_{12}$ well above the Néel temperature $T_N$=7.4 K. Diffuse neutron scattering also demonstrates the presence of the short-range order reflexes in $HoB_{12}$ at temperatures at least up to $3T_N$ [21]. The energy shift of the CF peaks below 30 K (see inset in Fig. 3) could be also attributed to the occurrence of weak internal inhomogeneous molecular field (MF) due to short-range magnetic order.

The influence of antiferromagnetic ordering on the excitation spectrum has been taken into account in the MF approximation with Zeeman energy $H_Z = \pm g_J \mu_B \mathbf{J} \cdot \mathbf{B}_{loc}$ ($\mathbf{B}_{loc}$ is the staggered local magnetic field affecting $Ho^{3+}$ ions in antiferromagnetically coupled dimers considered below) in the right-hand side in Eq. (1) when considering the magnetically ordered phase.

The molecular field in the antiferromagnetic state has been determined directly by fitting the INS spectra measured on NERA and DIN 2PI spectrometers at 3.5 K, assuming that the CF parameters are the same as in the paramagnetic phase. The best fit is obtained with $B_{loc}$ = (1.75± 0.1) T, and it can be seen in Fig. 3 that the CF + MF model describes very well the INS experimental data at both instruments. The measured and simulated values of excitation energies at 3.5 K are given in Table I.

The CF parameters in $HoB_{12}$, determined in the present work, have the same signs as the ones determined in $Yb_{0.9}Er_{0.1}B_{12}$ [30] for $Er^{+3}$ ions ($B_4$=-1.036±0.721 meV and $B_6$=-1.932 meV) and in $Lu_{1-x}Tm_xB_{12}$ and $Yb_{1-x}Tm_xB_{12}$ for $Tm^{+3}$ ions ($B_4$= -0.597 meV and $B_6$=-2.163 meV) [31], but, contrary to the parameter $B_6$ that demonstrates rather weak relative changes along the lanthanide series in $RB_{12}$, the absolute value of the parameter $B_4$ increases remarkably with the increasing number of the 4*f* electrons. We would like to note a specific feature of the sets of CF parameters in dodecaborides with unconventional large ratios $B_6/B_4$ which may evidence for dominant contributions into the CF potential from conduction electrons with a *f*-type space distribution (corresponding to the orbital moment *l*=3) in the vicinity of RE ions.

The present study demonstrates the importance of accounting for *J*-mixing effects due to spin-orbit and CF interactions even in the case of well isolated multiplet of a RE ion, especially, for estimations CF parameters with small absolute values. In particular, we found it possible to obtain practically the same CF energies and INS profiles using the three times diminished parameter $B_4$ (but only an eight percents smaller parameter $B_6$) in the CF Hamiltonian defined in the truncated basis of states of the $^5I_8$ multiplet that seems doubtful.

A possible reason for specific features of the CF parameters and their dependence on RE ion in borides may be the local polarization of conduction electrons and, as a consequence, a change of the conduction electrons contribution into the CF potential [32]. We plan to conduct systematic high precision measurements of the CF parameters along the series of $RB_{12}$ compounds in order to verify this assumption.

## V. MAGNETIZATION AND HEAT CAPACITY RESULTS AND ANALYSIS

The calculated value of the *g*-factor in Spin Hamiltonian of the $\Gamma_5^{(1)}$ ground triplet $g=5.56$ in absence of the orbital reduction ($k=1$) exceeds remarkably the *g*-factors measured in EPR experiments [29] varying with temperature from 4.8 (30 K) to 4.1 (7.5 K). Also, a substantial systematic decrease of the magnetization was unveiled by the magnetometry of the concentration series of isostructural $Ho_xLu_{1-x}B_{12}$ single crystals with the increasing concentration of $Ho^{3+}$ ions $x=0.01, 0.04, 0.1, 0.2. 03, 0.5, 1$ (not shown) [33]. We consider here both a single-ion and collective mechanism of the magnetic moment reduction in RE dodecaborides. First of all, the orbital magnetic moment of a localized 4*f* electron is reduced due to the hybridization of its wave functions with the wave functions of conduction electrons. Additional reduction of the effective magnetic moment can be induced by low-symmetry components of the crystal-field generated by random lattice strains which compete with the magnetic field because of different symmetry relative to the time inversion. This effect is important at low temperatures when energies $k_BT$ of thermal excitations are comparable to random splittings of the degenerate ground state of a RE ion [34]. The hybridization of wave functions can be accounted for by introduction of the orbital reduction factor $k<1$. We found it possible to reproduce successfully the measured field dependences of magnetization ***M(B)*** of the diluted system $Ho_{0.01}Lu_{0.99}B_{12}$ in magnetic fields up to 18 T directed along the [111], [110] and [001] axes at 8 K in the framework of the CF model derived above by making use of the orbital reduction factor $k=0.91$ (see Fig. 4a). Note, the CF model is in agreement with the observed cubic magnetic anisotropy [21], $M(\boldsymbol{B}\|[111])\approx M(\boldsymbol{B}\|[110])>M(\boldsymbol{B}\|[001])$, that becomes pronounced in the fields $B>3$ T.

In magnetically concentrated system $HoB_{12}$, magnetic interactions between $Ho^{3+}$ ions play the dominant role in the suppression of effective magnetic moments and formation of specific features of the magnetization field and temperature dependencies, such as the linear dependence on the field strength and no signs of the saturation or magnetic anisotropy in the fields up to 4.5 T (see Fig. 4b). The concrete magnetic structure in $HoB_{12}$ at temperatures below $T_N$ is still unknown (see discussion in [21]), and to account explicitly for the short-range antiferromagnetic correlations, we consider a minimal cluster model, an exchange-coupled holmium dimer in self-consistent molecular field $\boldsymbol{B}_{MF} = -\lambda <\boldsymbol{M}>$ (here $<\boldsymbol{M}>$ is the average holmium magnetic moment in external magnetic field in the paramagnetic phase and $\lambda$ is the molecular field constant linear in the corresponding exchange integral) originating from inter-dimer interactions. Note, a strong reduction of the magnetization of RE dimers allocated by the ladder space structure was revealed in RE higher borides $RB_{50}$ (see [35,36] and references therein). The Hamiltonian of a dimer determined in the space of $17 \times 17$ states corresponding to the CF sublevels of the ground multiplets of a pair of holmium ions has the form

$$H_D = H_1 + H_2 - J_{ex}\boldsymbol{S}_1 \cdot \boldsymbol{S}_2 - (\boldsymbol{M}_1 + \boldsymbol{M}_2) \cdot \boldsymbol{B}_{MF} \tag{4}$$

where $\boldsymbol{S}_i$ and $\boldsymbol{M}_i$ ($i$=1,2) are operators of the total spin moment and magnetic moment of a $Ho^{3+}$ ion, and $H_i$ is the single-ion Hamiltonian (1) projected on the wave functions of the $^5I_8$ multiplet of the $4f^{10}$ configuration. The exchange integral $J_{ex}$, as well as the mean field constant $\lambda$, are considered as the fitting parameters. The average values of the magnetic moments $<\boldsymbol{M}(\boldsymbol{B})>$ are solutions of the coupled self-consistent equations

$$<\boldsymbol{M}> = \frac{\text{Trace}[(\boldsymbol{M}_1 + \boldsymbol{M}_2)\exp(-H_D(\boldsymbol{B}_{loc})/k_BT)]}{2\text{Trace}[\exp(-H_D(\boldsymbol{B}_{loc})/k_BT)]}, \tag{5}$$

$$<\boldsymbol{M}> = (\boldsymbol{B}_{loc} - \boldsymbol{B})/\lambda, \tag{6}$$

where $\boldsymbol{B}_{loc} = \boldsymbol{B} + \boldsymbol{B}_{MF}$. From fitting of the magnetization curve at the temperature 7.4 K (see Fig. 4b), we obtained $J_{ex}$=-2.55 K and $\lambda$=-0.38 T/$\mu_B$ (negative signs correspond to antiferromagnetic interactions), but, to reproduce the experimental data at higher temperatures, we found it necessary to decrease slightly the exchange integral and, correspondingly, the parameter $J_{ex}$=-2.39 K ($T$=10 K), -2.23 K ($T$=20 K), -2.09 K ($T$=30 K) and -1.92 K ($T$=50 K). The results of modelling match satisfactory the measured field dependencies of the holmium magnetic moments in the paramagnetic phase. Changes of the exchange integrals may be caused by the thermal lattice expansion, or by decreasing the Ho-Ho distances in dimers with temperature lowering in the disordered cage-glass phase.

It should be noted that the calculation of profiles of INS by noninteracting dimers by making use the CF and exchange parameters presented above brings about practically the same results as in case of noninteracting holmium ions in the paramagnetic phase because the exchange splitting of degenerate CF sublevels does not exceed the spectral resolution of 0.8 meV.

Next, the derived models were checked by calorimetric studies of $HoB_{12}$ single crystals in the paramagnetic phase in external magnetic fields. Figure 5a shows the measured temperature dependencies of heat capacity $C(T, B)$ in magnetic fields $B$=0, 2, 4, 6 and 8 T, Einstein component $C_E$, phonon contribution $C_{ph}$ and magnetic specific heat $C_m(T, B=0)$ where the last term is separated from phonon and conduction electron components by subtracting the specific heat of the non-magnetic reference LuB$_{12}$ compound (see Ref. [34] for more details). Additionally, we took into account the difference in Einstein temperatures of HoB$_{12}$ ($\theta_E \approx 190$ K [37]) and LuB$_{12}$ ($\theta_E \approx 164$ K) [1,3,34] using the relation

$$C_m(\text{HoB}_{12}) = [C(\text{HoB}_{12}) - C_E(\text{HoB}_{12})] - [C(\text{LuB}_{12}) - C_E(\text{LuB}_{12})] \qquad (7)$$

and the well-known Einstein formula for the specific heat [1]

$$C_E = \sum_j R \left(\frac{T_j}{T}\right)^2 \frac{e^{T_j/T}}{\left(e^{T_j/T} - 1\right)^2} \qquad (8)$$

where $R$ is the gas constant, $T_j = \hbar \omega_j / k_B$, and the sum is taken over Einstein phonons with frequencies $\omega_j$. The temperature dependencies of $C_m(T, B)$ obtained in accordance with Eq. (7) are shown in Fig. 5b. Near Néel temperature $T_N(B) \leq 7.4$ K, curves $C_m(T, B)$ demonstrate a stepwise rise [20,21] (not shown in Fig. 5) whose amplitude steeply decreases with the magnetic field. In the paramagnetic state the $C_m(T, B)$ plots exhibit a broad peak centered at about 50 K (Fig. 5b), which corresponds to the magnetic contribution of the holmium subsystem from the excited CF levels. At low temperatures an additional maximum appears on $C_m(T, B)$ curves which may be attributed to a Zeeman-type Schottky anomaly in the heat capacity. To shed more light on the magnetic field effect on heat capacity, field induced differences

$$\Delta C(T, B) = C(T, B) - C(T, B=0) \qquad (9)$$

were introduced, and the corresponding plots are presented in Fig. 6a. The temperature dependencies of $\Delta C(T, B)$ demonstrate a wide maximum at about $T_{\max}^C \approx 12$ K, the maximum amplitude increases strongly with magnetic field (see inset in Fig. 6a). Only very weak decrease of $T_{\max}^C(B)$ observed in Fig. 6a agrees qualitatively with the marked above reduction of holmium magnetic moments and, correspondingly, of Zeeman splittings of CF levels (see also inset in Fig. 3). Short-range order effects playing important role in this reduction were also detected in HoB$_{12}$ in Refs. [19-21,29,38,39].

We would like to notice that the analysis of field induced differences $\Delta C(T, B)$ allowed us to detect an additional fine effect in the heat capacity of HoB$_{12}$. Indeed, a large scale presentation of $\Delta C(T, B)$ in the range 40-80 K demonstrates the cage-glass transition at $T^* \sim 60$ K which is accompanied by a step-like anomaly for **B** // [001] and **B** // [110], i. e., in magnetic fields transverse to the *cigar-shaped AFM-correlated regions* [21], but such a singularity is absent in fields along the [111] direction (see Fig. 6b).

The calculated temperature dependence of the molar heat capacity of the subsystem of noninteracting $Ho^{3+}$ ions in the cubic crystal field (the lowest dashed line in Fig. 5b) reproduces satisfactory the zero-field experimental curve $C_m(T, 0)$ in the range 30-150 K, but at low temperatures the results of calculations are remarkably underestimated because of neglecting random lattice strains and low-energy excitations with a continuous spectrum originating from magnetic dipole and exchange interactions. In the framework of the dimer model introduced above, the heat capacity is determined by the expression

$$C_m(T,B) = \frac{N_A}{2k_B T^2}[<H_D^2> - <H_D>^2] \qquad (10)$$

where $N_A$ is the Avogadro number, angular brackets denote the quantum statistical averaging over the canonical ensemble, and $H_D$ is the dimer Hamiltonian (4). The results of calculations for $B$=0 (see line and symbol curve in Fig. 5b) where we have employed the CF parameters and the exchange integral obtained from modelling the INS spectra and the low temperature magnetization agree substantially better with the experimental curve, but remain underestimated.

The detailed theoretical analysis of low-energy excitations in $HoB_{12}$ is out of scopes of the present work. To describe the low temperature behavior of heat capacity, we used the phenomenological approach keeping the dimer model but, having in mind the increasing role of antiferromagnetic interactions in external magnetic fields which align magnetic moments of neighboring $Ho^{3+}$ ions, with varying (increasing with magnetic field) values of the exchange integral and the corresponding mean field constant $\lambda$. The satisfactory fitting was achieved by making use the measured average magnetic moments $<M(T,B)>$ of $Ho^{3+}$ ions (see Fig. 4b) and the exchange integrals $J_{ex}$=3.0 K ($B$=0 and 2 T), 3.9 K ($B$=4 T), 4.3 K ($B$=6 T) and 4.4 K ($B$=8 T). The calculated differences $\Delta C_m(T,B) = C_m(T,B) - C_m(T,0)$ are compared with the measured ones $\Delta C(T,B)$ for $B$=2, 4, 6 and 8 T in Fig. 6a. The systematic discrepancies between the results of modeling and the experimental data at temperatures below ~12 K point to a possible temperature dependence (enhances with decreasing temperature) of the exchange integral in the fixed external magnetic field. Taking into account the loosely bound state of $Ho^{3+}$ ions in the rigid covalent boron network of $HoB_{12}$ and the appearance of dynamic charge stripes along the [110] directions below $T^* \sim 60$ K, it could be assumed that a significant decrease in the Ho-Ho distance in dimers is induced by a strong external magnetic field, and this assumption should be tested in future experiments.

## VI. CONCLUSION

To summarize, the strongly correlated electron system Ho$^{11}$B$_{12}$ with Jahn-Teller instability of rigid boron network and dynamic charge stripes was studied in detail by inelastic neutron scattering, magnetometry and heat capacity measurements at temperatures in the range 3-300 K.

The CF parameters $B_4$=- 0.333 meV and $B_6$= -2.003 meV for Ho$^{3+}$ ions in the cubic crystal field were obtained from the analysis of INS spectra, and the absolute value of $B_4$ turned out to be significantly lower than the values determined earlier in other magnetic RE (erbium and thulium) dodecaborides. The large ratio $B_6/B_4$ indicates probably the dominant contribution of conduction electrons into the CF potential. At temperatures below 30 K, short-range order effects have been found in INS spectra in combination with specific field and temperature dependencies of the low-temperature magnetization and heat capacity which may be discussed in terms of formation below $T^*$~60 K of holmium dimers in the paramagnetic state of HoB$_{12}$ with the cage-glass structure. The molecular field in the antiferromagnetic state, $B_{MF}$ = (1.75± 0.1) T has been determined by fitting the INS spectra at 3.5 K, assuming that the CF parameters are the same as in the paramagnetic phase.

The determination of the magnetic structure of HoB$_{12}$ in the magnetically ordered phase remains a challenging problem.

**Acknowledgments**

This work in Prokhorov General Physics Institute of RAS was supported by the Russian Science Foundation, Project No. 17-12-01426 and Russian Foundation for Basic Research, Project No. 18-02-01152 and partly performed using the equipment of the Center of Excellence, Slovak Academy of Sciences. The work of K.F., S.G., M.R. and K.S. was supported by the projects APVV-17-0020 and DAAD-57561069. B.Z.M. acknowledges the support by the Russian Science Foundation, Project No. 19-12-00244. The authors are grateful to S.V. Demishev and V.V. Glushkov for helpful discussions.

**References**

[1] N. E. Sluchanko, A. N. Azarevich, A. V. Bogach, I. I. Vlasov, V. V. Glushkov, S. V. Demishev, A. A. Maksimov, I. I. Tartakovskii, E. V. Filatov, K. Flachbart, S. Gabani, V. B. Filippov, N. Y. Shitsevalova, and V. V. Moshchalkov, Effects of disorder and isotopic substitution in the specific heat and Raman scattering in LuB$_{12}$, JETP **113**, 468 (2011).

[2] N. E. Sluchanko, A. V. Bogach, V. V. Glushkov, S. V. Demishev, K. S. Lyubshov, D. N. Sluchanko, A. V. Levchenko, A. B. Dukhnenko, V. B. Filipov, S. Gabani, and K. Flachbart,

Antiferromagnetic instability and the metal-insulator transition in $Tm_{1-x}Yb_xB_{12}$ rare earth dodecaborides, JETP Lett. **89**, 256 (2009).

[3] A. Czopnik, N. Shitsevalova, A. Krivchikov, V. Pluzhnikov, Y. Paderno, and Y. Onuki, Thermal properties of rare earth dodecaborides, J. Solid State Chem. **177**, 507 (2004).

[4] F. Iga, N. Shimizu, and T. Takabatake, Single crystal growth and physical properties of Kondo insulator $YbB_{12}$, J. Magn. Magn. Mater. **177**, 337 (1998).

[5] F. Iga, Y. Takakuwa, T. Takahashi, M. Kasaya, T. Kasuya, and T. Sagawa, XPS Study of Rare-Earth Dodecaborides - $TmB_{12}$, $YbB_{12}$ and $LuB_{12}$, Solid State Commun. **50**, 903 (1984).

[6] T. S. Altshuler, A. E. Altshuler, and M. S. Bresler, An EPR study of the temperature dependence of the energy gap in ytterbium dodecaboride, JETP **93**, 111 (2001).

[7] K. Hagiwara, Y. Takeno, Y. Ohtsubo, R. Yukawa, M. Kobayashi, K. Horiba, H. Kumigashira, J. Rault, P. Le Fèvre, and F. Bertran, Temperature dependence of Yb valence in the sub-surface of $YbB_{12}$(001), J. Phys.: Conf. Ser. **807**, 012003 (2017).

[8] J. Otsuki, H. Kusunose, P. Werner, and Y. Kuramoto, Continuous-time quantum Monte Carlo method for the Coqblin-Schrieffer model, J. Phys. Soc. Jpn. **76**, 114707 (2007).

[9] A. Akbari, P. Thalmeier, and P. Fulde, Theory of Spin Exciton in the Kondo Semiconductor $YbB_{12}$, Phys. Rev. Lett. **102**, 106402 (2009).

[10] A. F. Barabanov and L. A. Maksimov, Spin excitations in Kondo insulator $YbB_{12}$, Phys. Lett. A **373**, 1787 (2009).

[11] F. Lu, J. Z. Zhao, H. M. Weng, Z. Fang, and X. Dai, Correlated Topological Insulators with Mixed Valence, Phys. Rev. Lett. **110**, 096401 (2013).

[12] H. M. Weng, J. Z. Zhao, Z. J. Wang, Z. Fang, and X. Dai, Topological Crystalline Kondo Insulator in Mixed Valence Ytterbium Borides, Phys. Rev. Lett. **112**, 016403 (2014).

[13] K. Hagiwara, Y. Ohtsubo, M. Matsunami, S. Ideta, K. Tanaka, H. Miyazaki, J. E. Rault, P. Le Fevre, F. Bertran, A. Taleb-Ibrahimi, R. Yukawa, M. Kobayashi, K. Horiba, H. Kumigashira, K. Sumida, T. Okuda, F. Iga, and S. Kimura, Surface Kondo effect and non-trivial metallic state of the Kondo insulator $YbB_{12}$, Nat. Commun. **7**, 12690 (2016).

[14] N. Sluchanko, A. Bogach, N. Bolotina, V. Glushkov, S. Demishev, A. Dudka, V. Krasnorussky, O. Khrykina, K. Krasikov, V. Mironov, V. B. Filipov, and N. Shitsevalova, Rattling mode and symmetry lowering resulting from the instability of the $B_{12}$ molecule in $LuB_{12}$, Phys. Rev. B **97**, 035150 (2018).

[15] N. B. Bolotina, A. P. Dudka, O. N. Khrykina, V. N. Krasnorussky, N. Y. Shitsevalova, V. B. Filipov, and N. E. Sluchanko, The lower symmetry electron-density distribution and the charge transport anisotropy in cubic dodecaboride $LuB_{12}$, J. Phys.: Condens. Matter **30**, 265402 (2018).

[16] N. B. Bolotina, A. P. Dudka, O. N. Khrykina, V. V. Glushkov, A. N. Azarevich, V. N. Krasnorussky, S. Gabani, N. Y. Shitsevalova, A. V. Dukhnenko, V. B. Filipov, and N. E. Sluchanko, On the role of isotopic composition in crystal structure, thermal and charge-transport characteristics of dodecaborides $Lu^NB_{12}$ with the Jahn-Teller instability, J. Phys. Chem. Sol. **129**, 434 (2019).

[17] N. E. Sluchanko, A. N. Azarevich, A. V. Bogach, N. B. Bolotina, V. V. Glushkov, S. V. Demishev, A. P. Dudka, O. N. Khrykina, V. B. Filipov, N. Y. Shitsevalova, G. A. Komandin, A. V. Muratov, Y. A. Aleshchenko, E. S. Zhukova, and B. P. Gorshunov, Observation of dynamic charge stripes in $Tm_{0.19}Yb_{0.81}B_{12}$ at the metal-insulator transition, J. Phys.: Condens. Matter **31**, 065604 (2019).

[18] B. P. Gorshunov, E. S. Zhukova, G. A. Komandin, V. I. Torgashev, A. V. Muratov, Y. A. Aleshchenko, S. V. Demishev, N. Y. Shitsevalova, V. B. Filipov, and N. E. Sluchanko, Collective Infrared Excitation in $LuB_{12}$ Cage-Glass, JEPT Lett. **107**, 100 (2018).

[19] A. L. Khoroshilov, V. N. Krasnorussky, K. M. Krasikov, A. V. Bogach, V. V. Glushkov, S. V. Demishev, N. A. Samarin, V. V. Voronov, N. Y. Shitsevalova, V. B. Filipov, S. Gabani, K. Flachbart, K. Siemensmeyer, S. Y. Gavrilkin, and N. E. Sluchanko, Maltese cross anisotropy in $Ho_{0.8}Lu_{0.2}B_{12}$ antiferromagnetic metal with dynamic charge stripes, Phys. Rev. B **99,** 174430 (2019).

[20] K. Krasikov, V. Glushkov, S. Demishev, A. Khoroshilov, A. Bogach, V. Voronov, N. Shitsevalova, V. Filipov, S. Gabáni, K. Flachbart, K. Siemensmeyer, and N. Sluchanko, Suppression of indirect exchange and symmetry breaking in the antiferromagnetic metal $HoB_{12}$ with dynamic charge stripes, Phys. Rev. B, **102**, 214435 (2020).

[21] K. Siemensmeyer, K. Habicht, T. Lonkai, S. Mat'as, S. Gabani, N. Shitsevalova, E. Wulf, and K. Flachbart, Magnetic properties of the frustrated fcc - Antiferromagnet $HoB_{12}$ above and below $T_N$, J. Low Temp. Phys. **146**, 581 (2007).

[22] I. Natkaniec, D. Chudoba, L. Hetmańczyk, V. Yu. Kazimirov, J. Krawczyk, I. L. Sashin Parameters of the NERA spectrometer for cold and thermal moderators of the IBR-2 pulsed reactor, J. Phys.: Conf. Ser. **554**, 012002 (2014).

[23] I. V. Kalinin, V. M. Morozov, A. G. Novikov, A. V. Puchkov, V. V. Savostin, V. V. Sudarev, A. P. Bulkin, S. I. Kalinin, V. M. Pusenkov, and V. A. Ul'yanov, Characteristics of the DIN-2PI spectrometer with a neutron concentrator, Technical Physics, **59**, 307 (2014).

[24] W. T. Carnall, G. L. Goodman, K. Rajnak, and R. S. Rana, A systematic analysis of the spectra of the lanthanides doped into single crystal $LaF_3$, J. Chem. Phys. **90,** 3443 (1989).

[25] D. S. Pytalev, E. P. Chukalina, M. N. Popova, G. S. Shakurov, B. Z. Malkin, S. L. Korableva, Hyperfine interactions of $Ho^{3+}$ ions in $KY_3F_{10}$: Electron paramagnetic resonance and optical spectroscopy studies, Phys. Rev. B **86**, 115124 (2012).

[26] A. Abragam and B. Bleaney, Electron Paramagnetic Resonance of Transition ions, Clarendon, Oxford (1970).

[27] P. Fulde and M. Loewenhaupt, Magnetic excitations in crystal-field split 4f systems, Adv. Phys. **34**, 589 (1985).

[28] M. A. H. McCausland and I. S. Mackenzie, Nuclear magnetic resonance in rare earth metals, Adv. Phys. **28,** 305 (1979).

[29] M. I. Gilmanov, S. V. Demishev, B. Z. Malkin, A. N. Samarin, N. Yu. Shitsevalova, V. B. Filipov, and N. E. Sluchanko, Electron paramagnetic resonance in $Ho_xLu_{1-x}B_{12}$ dodecaborides, JETP Lett. **110**, 241 (2019).

[30] P. A. Alekseev, J.-M. Mignot, K. S. Nemkovski, E. V. Nefeodova, N. Yu. Shitsevalova, Yu. B. Paderno, R. I. Bewley, R. S. Eccleston, E. S. Clementyev, V. N. Lazukov, I. P. Sadikov and N. N. Tiden, Yb–Yb correlations and crystal-field effects in the Kondo insulator $YbB_{12}$ and its solid solutions, J. Phys.: Condens. Matter **16**, 2631 (2004).

[31] P. A. Alekseev, K. S. Nemkovski, J.-M. Mignot, E. S. Clementyev, A. S. Ivanov, S. Rols, R. I. Bewley, V. B. Filipov, and N. Yu. Shitsevalova, Possible undercompensation effect in the Kondo insulator $(Yb,Tm)B_{12}$, Phys. Rev. B **89**, 115121 (2014).

[32] E. A. Goremychkin and E. Myule, Conduction-electron contribution to the crystal potential in intermetallic compounds of rare-earth metals, JETP Lett. **39**, 570 (1984).

[33] N. E. Sluchanko et al., to be published.

[34] N. E. Sluchanko, A. L. Khoroshilov, A. V. Bogach, S. Yu. Gavrilkin, V. V. Glushkov, S. V. Demishev, V. N. Krasnorussky, N. Yu. Shitsevalova, V. B. Filipov, S. Gabani, K. Flachbart, B. Z. Malkin, Magnetic Anisotropy of the Low-Temperature Specific Heat of $Ho_{0.01}Lu_{0.99}B_{12}$ with Dynamic Charge Stripes, JETP Lett. **108**, 454 (2018).

[35] S. Gabani, K. Flachbart, K. Siemensmeyer and T. Mori, Magnetism and superconductivity of rare earth borides, J. Alloy. Comp. **821**, 153201 (2020).

[36] B. Z. Malkin, S. L. Bud'ko, and V. V. Novikov, Crystal-field approach to rare-earth higher borides: Dimerization, thermal, and magnetic properties of $RB_{50}$ (R = Tb, Dy, Ho, Er, Tm), Phys. Rev. Mater. **4**, 054409 (2020).

[37] N. B. Bolotina (private commun.).

[38] K. M. Krasikov, A. V. Bogach, A. D. Bozhko, V. V. Glushkov, S. V. Demishev, A. L. Khoroshilov, N. Yu. Shitsevalova, V. Filipov, S. Gabáni, K. Flachbart, and N. E. Sluchanko, Anisotropy of the charge transport in $Ho^{11}B_{12}$ antiferromagnet with dynamic charge stripes, Solid State Sci. **104**, 106253 (2020).

[39] N. E. Sluchanko, Magnetism, quantum criticality, and metal-insulator transitions in $RB_{12}$, in *Rare-Earth Borides*, edited by D. S. Inosov (Jenny Stanford Publishing, Singapore, 2021), Chap. 4, pp. 331-442, ISBN: 978-981-4877-56-5, 978-1-003-14648-3 (eBook); www.jennystanford.com.

Table I. Measured and calculated energies (meV) of CF sublevels of the ground multiplet $^5I_8$ of Ho$^{3+}$ ions in Ho$^{11}$B$_{12}$ and the widths $\Delta_{HF}$ (meV) of hyperfine splittings of corresponding sublevels.

| Symmetry and degeneracy | $E_i$ ($T>T_N$) Experiment | $E_i$ ($T>T_N$) Theory | $\Delta_{HF}$ Theory | CF excitations in INS at 3.5 K Experiment | CF excitations in INS at 3.5 K Theory |
|---|---|---|---|---|---|
| $\Gamma_5^{(1)}$  3 | 0 | 0 | 0.216 | 0 and 0.8 | 0 and 0.83 |
| $\Gamma_3^{(1)}$  2 | 7.2 | 7.18 | 0.005 | 8.0 | 7.94 |
| $\Gamma_4^{(1)}$  3 | 9.5 | 9.52 | 0.017 | 11.0 | 10.39 |
| $\Gamma_1$  1 | - | 12.30 | 0.004 | - | - |
| $\Gamma_5^{(2)}$  3 | 17.4 | 17.28 | 0.192 | 18.5 | 18.69 |
| $\Gamma_4^{(2)}$  3 | 26.7 | 26.58 | 0.152 | 26.8 | 26.62 |
| $\Gamma_3^{(2)}$  2 | - | 27.05 | 0.042 | - | - |

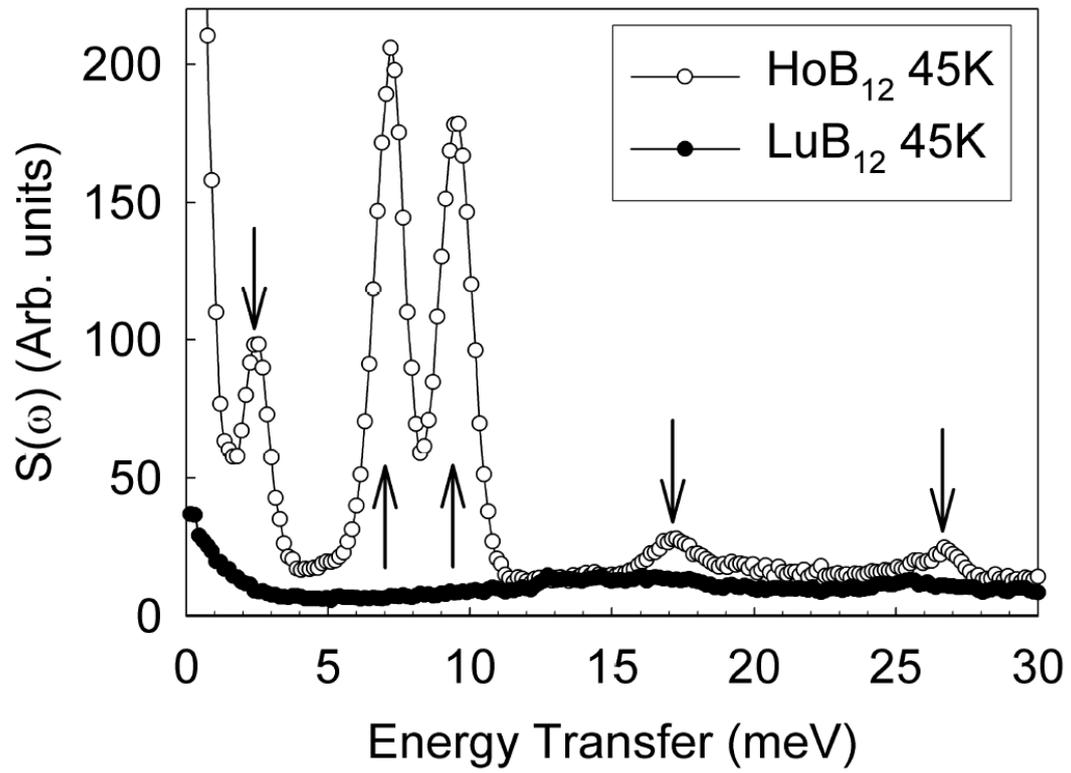

FIG. 1. Inelastic neutron scattering spectra of Ho$^{11}$B$_{12}$ and Lu$^{11}$B$_{12}$ measured at 45 K on NERA. The arrows show the positions of the peaks corresponding to CF excitations.

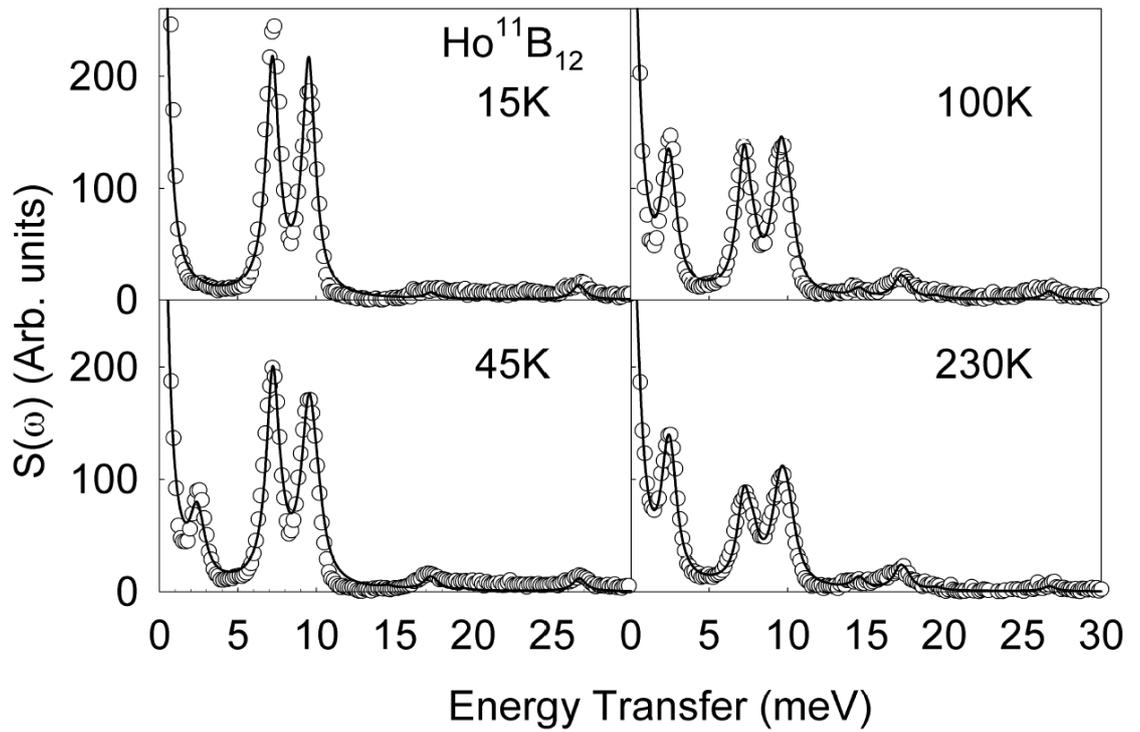

FIG. 2. Background corrected inelastic neutron scattering spectra from Ho$^{11}$B$_{12}$ measured at 15 K, 45 K, 100 K and 230 K on NERA. The solid lines show the profiles calculated in the framework of the CF model described in the text.

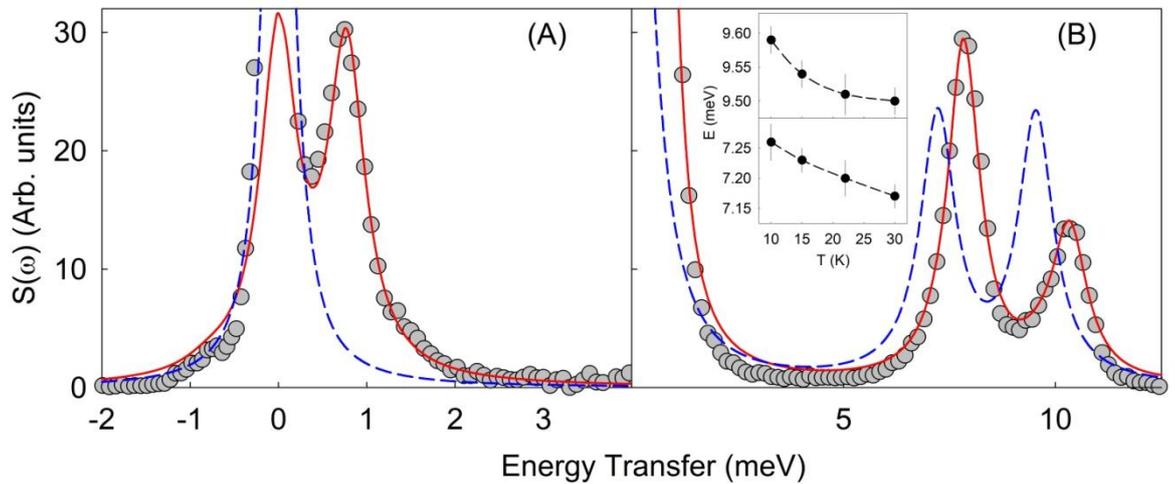

FIG. 3. Background corrected inelastic neutron scattering spectra of Ho$^{11}$B$_{12}$ measured at 3.5 K for energies of incident neutrons of 5 meV on DIN 2PI (A) and for registered energies of 4.65 meV on NERA (B). The red solid line shows the profile of the CF + MF model described in the text. The blue dashed line represents the CF spectrum only. Inset shows the energy shift of the CF peaks below 30 K.

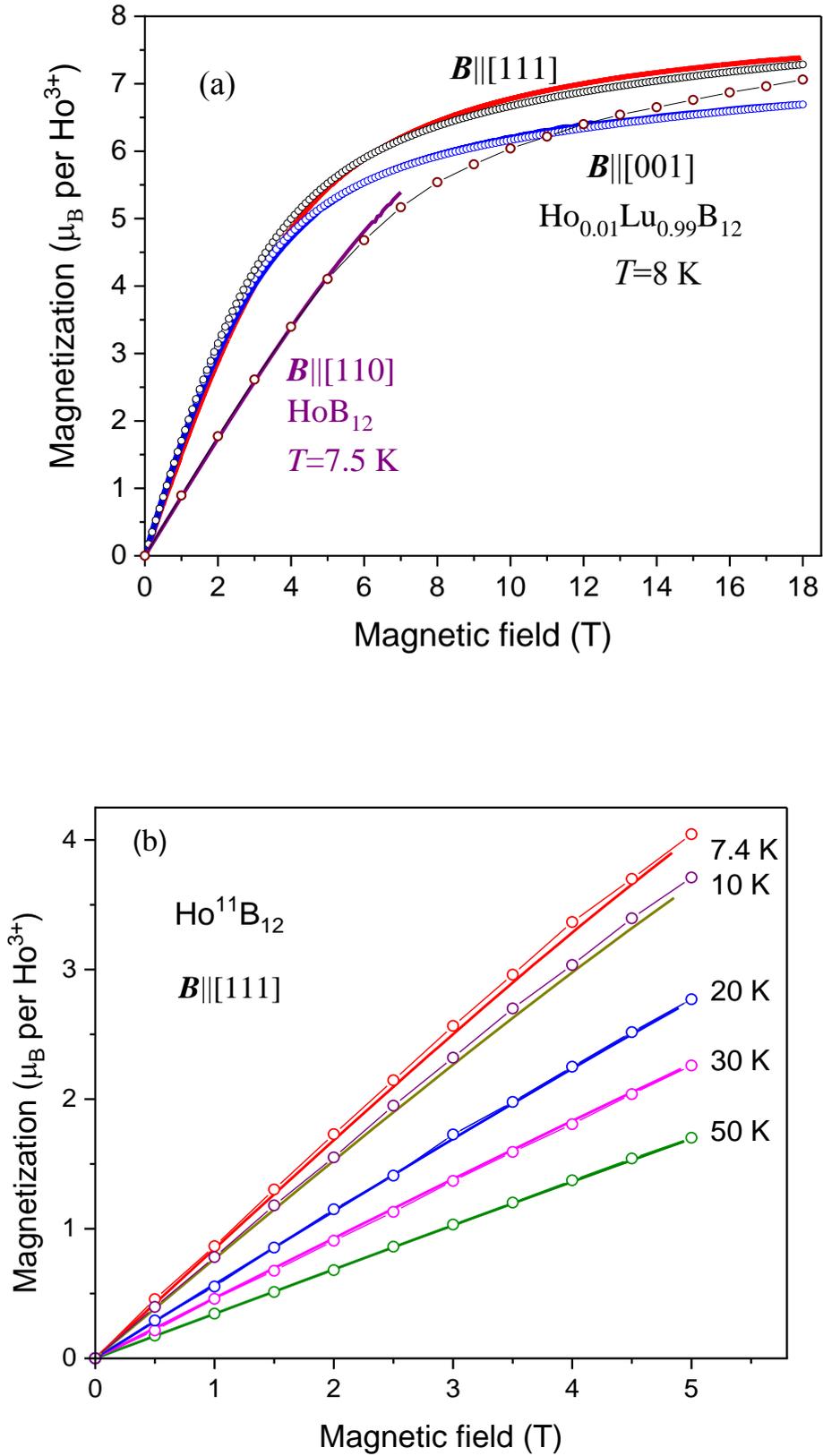

FIG. 4. Measured (solid lines) and calculated (symbols) magnetization of (a) $Ho_{0.01}Lu_{0.99}B_{12}$ (for a direct comparison, the measured magnetization of the concentrated system $HoB_{12}$ is also shown) and (b) $Ho^{11}B_{12}$ single crystals in magnetic fields $B \parallel [111]$ at temperatures 7.4, 10, 20, 30 and 50 K.

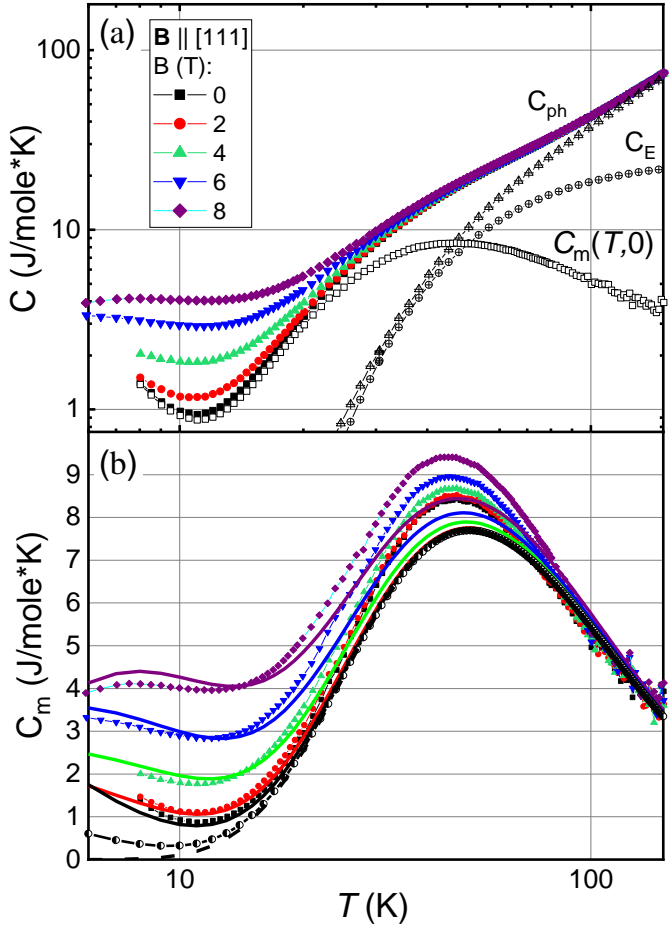

FIG. 5. (a) Temperature dependencies of specific heat $C(T, B)$ for $\mathbf{B} \parallel [111]$, Einstein component $C_E$, phonon contribution $C_{ph}$ and magnetic heat capacity $C_m(T,0)$ of $HoB_{12}$. (b) The results of $C_m(T, B)$ modeling. Thick solid lines represent the fitting in magnetic fields $B$=0 T, 2 T, 4 T, 6 T and 8 T in the framework of a dimer model. The lowest dashed line and the line with circles show the calculated heat capacity in zero magnetic field for noninteracting $Ho^{3+}$ ions and holmium dimers (with the intra-dimer exchange integral $J_{ex}$=-2.55 K determined from the analysis of the magnetization), respectively.

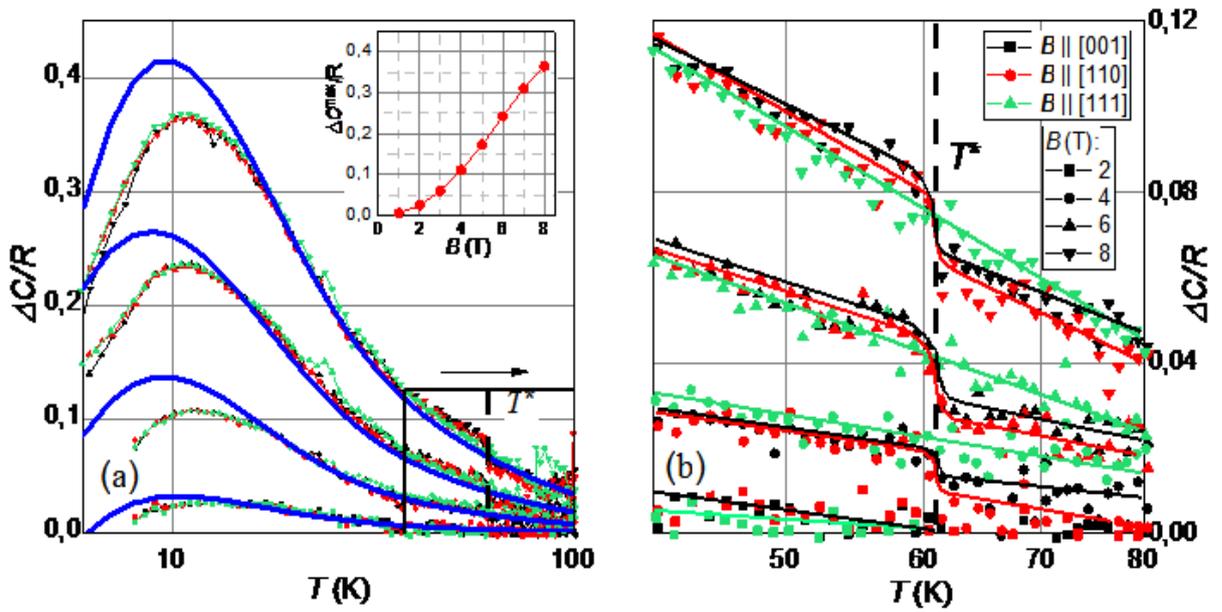

FIG. 6. (a) Measured temperature dependences of the magnetic field induced difference $\Delta C=C(T, B)-C(T, 0)$ for $B \parallel [001]$, $B \parallel [110]$ and $B \parallel [111]$ configurations (symbols) in $HoB_{12}$. Solid lines correspond to the results of simulations. Inset shows the magnetic field dependence of the $\Delta C$ maximum. (b) Large scale plot of $\Delta C(T, T)$ in the temperature range 40-80 K. $T^*$ denotes the cage-glass transition anomaly.